\documentclass{article}
\usepackage{hq2kpro}
\begin{document}
%
\def \ie    {\hbox{\it i.e.}}     
\def \etc   {\hbox{\it etc.}}
\def \vs    {\hbox{\it vs.}}
\def \eg    {\hbox{\it e.g.}}     
\def \etal  {\hbox{\it et al.}}
\def \Journal#1#2#3#4{{#1} {\bf #2}, #3 (#4)}
\def \CPC   {Comput. Phys. Commun.}
\def \EPJ   {European Phys. Journal}
\def \NIM   {Nucl. Instr. Meth.}
\def \NP    {Nucl. Phys.}
\def \PL    {Phys. Lett.}
\def \PR    {Phys. Rev.}
\def \PRL   {Phys. Rev. Lett.}
\def \PRep  {Phys. Reports}
\def \RMP   {Revs. Mod. Phys.}
\def \zphys {Z. Phys.}
\hyphenation{back-ground}
\hyphenation{brem-sstrah-lung}
\hyphenation{cal-or-ime-ter cal-or-ime-try}
\hyphenation{had-ron had-ronic}
\hyphenation{like-li-hood}
\hyphenation{posi-tron posi-trons}
\hyphenation{semi-lep-tonic}
\hyphenation{syn-chro-tron}
\hyphenation{system-atic}
\def \beq   {\begin{equation}}
\def \eeq   {\end{equation}}
\def \bbeq  {\begin{eqnarray*}}
\def \ebeq  {\end{eqnarray*}}
\def \bbeqn  {\begin{eqnarray}}
\def \ebeqn  {\end{eqnarray}}
\def \Tr    {\mathop{\mathrm Tr}}
\def \Im    {\mathop{\mathrm Im}}
\def \Re    {\mathop{\mathrm Re}}
\def \omg#1 {\mbox {${\mathcal O}(#1)$}}
\def \avg#1 {$\left\langle #1\right\rangle$}
\def \to    {\rightarrow}
\def \bra#1 {$\left\langle #1\right|$}
\def \ket#1 {$\left| #1\right\rangle$}
\def \braket#1#2 {\left\langle #1\right. \left| #2\right\rangle}
\def \amp#1 {${\mathcalA}(#1)$}
\def \apgt  {\raisebox{-0.6ex}{$\stackrel{>}{\sim}$}}
\def \aplt  {\raisebox{-0.6ex}{$\stackrel{<}{\sim}$}}
\def \pma#1#2 {\mbox{\raisebox{-0.6ex}
           {$\stackrel{\scriptstyle \;+\; #1}{\scriptstyle \;-\; #2}$}}}
\def \mev   {\,\mathrm {MeV}}
\def \gev   {\,\mathrm {GeV}}
\def \km    {\,\mathrm {km}}
\def \mm    {\,\mathrm {mm}}
\def \um    {\,\mu\mathrm m}
\def \ghz   {\,\mathrm {GHz}}
\def \mhz   {\,\mathrm {MHz}}
\def \khz   {\,\mathrm {kHz}}
\def \picos {\,\mathrm {ps}}
\def \ns    {\,\mathrm {ns}}
\def \us    {\,\mu\mathrm s}
\def \ms    {\,\mathrm {ms}}
\def \pb    {\,\mathrm {pb}$^{-1}$}
\def \fb    {\,\mathrm {fb}$^{-1}$}
\def \mrad  {\,\mathrm {mrad}}
\def \BR#1#2 {\mbox{$\mathcal Br$}(#1$\to$#2)}
\def \JP     {$\mathrm J^{\mathrm P}$}
\def \Mw    {${\mathrm M}_W$}
\def \Mpi   {${\mathrm M}_\pi$}
\def \Mk    {${\mathrm M}_K$}
\def \Gf    {G$_{\mathrm F}$}
\def \Mt    {${\mathrm M}_{t}$}
\def \Mb    {${\mathrm M}_{b}$}
\def \Mc    {${\mathrm M}_{c}$}
\def \Ms    {${\mathrm M}_{s}$}
\def \Mud   {${\mathrm M}_{u,d}$}
\def \lms   {\Lambda_{\overline{\mathrm MS}}}
\def \Vub   {$|$V$_{\mathrm{ub}}|$}
\def \Vus   {$|$V$_{\mathrm{ub}}|$}
\def \Vcb   {$|$V$_{\mathrm{cb}}|$}
\def \rphi {$r$-$\phi$}
\def \Pt2  {${\mathrm P}_\perp^2$}
\def \epem  {$e^+e^-$}
\def \mpmm  {$\mu^+\mu^-$}
\def \nunbr {$\nu \bar{\nu}$}
\def \lplm  {$\ell^+ \ell^-$}
\def \Pizd  {\mbox{$\pi^0_D$}}
\def \Kz    {\mbox{$K^0$}}
\def \Kzb   {\mbox{$\bar{K^0}$}}
\def \Kl    {\mbox{$K^0_L$}}
\def \Ks    {\mbox{$K^0_S$}}
\def \Kodd  {\mbox{$K_{\mathrm ODD}$}}
\def \Keven {\mbox{$K_{\mathrm EVEN}$}}
\def \Kp    {\mbox{$K^+$}}
\def \row   {\mbox{$\rho_{\mathrm CKM}$}}
\def \ate   {\mbox{$\eta_{\mathrm CKM}$}}
\def \Kggv  {\mbox{$K_L \gamma^{(*)} \gamma^{(*)}$}}
\def \akst  {\mbox{$\alpha_{K^*}$}}
%
%
%
%
\title{ 
Rare Decay Results from KTeV and (\row,~\ate)
}
\author{
Leo Bellantoni                              \\
{\em Fermi National Accelerator Laboratory} \\
for The KTeV Collaboration                  \\
}
\maketitle
\baselineskip=11.6pt
\begin{abstract}
Rare decay results from KTeV are reviewed, emphasizing modes that in
principle provide information about the CKM matrix.  Our recent results
in lepton flavor violating modes are also presented.
\end{abstract}
\baselineskip=14pt
\section{The Data Sample}

The KTeV results shown here are from the 1997 data sample, which
consists of about $2.7 \times 10^{11}$\ \Kl\ decay samples.  KTeV also
took data in 1999, giving a total data sample about 2.5 times what is
presented here for three body decays and about 3.2 times what is
presented here for four body decays.  A summary of the KTeV detector
is in the Appendix.

\section{Modes Relevant for \row}

\begin{figure}
\vspace{10.0cm}
\includegraphics{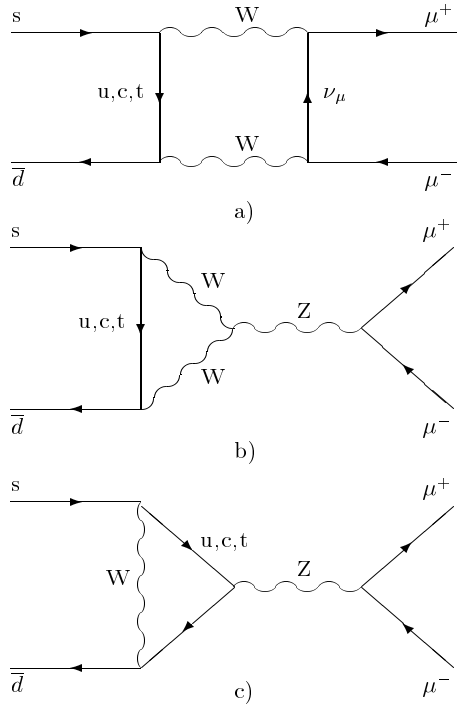}
\caption{\it
Interesting short distance contributions to \Kl $\to$ \mpmm.
\label{fig:mm_short} }
\end{figure}

\begin{figure}
\vspace{4.0cm}
\includegraphics{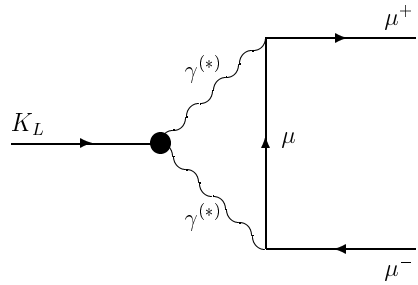}
\caption{\it
Boring long distance contributions to \Kl $\to $\mpmm.
\label{fig:mm_long} }
\end{figure}

The decay \Kl $\to$ \mpmm\ contains a short-distance contribution which
depend on the $\rho$\ parameter of the Wolfenstein formulation of
the CKM matrix.  The appropriate diagrams, shown in
Fig. \ref{fig:mm_short}, correspond to a branching ratio contribution
of\cite{ref:BnB1}

\beq
{\mathcal Br}(K^0_L \to \mu^+\mu^-) = 
\frac{\alpha^2{\mathcal Br}(K^+ \to \mu^+\nu)}{\pi^2 \sin^4 \theta_W}
\frac{\tau(K^0_L)}{\tau(K^+)}
\{(1-\frac{\lambda^2}{2})Y_{NL} + A^2\lambda^4(1-\rho)Y_t\}^2.
\label{eq:br}
\eeq

where $Y_{NL}$\ and $Y_t$\ incorporate next-to-leading order QCD
corrections, and $A$\ and $\lambda$\ are from the Wolfenstein
parameterization of the CKM matrix.  The important thing to notice here
is the dependence on \row.  Unfortunately, there are sizeable
long-range contributions, as shown in Fig. \ref{fig:mm_long}, as well.

The long distance contribution when both photons are real can be
determined with a measurement of \Kl $\to \gamma \gamma$\ and a pair of
QED vertices.\cite{ref:oldie,ref:2real}  The branching ratio for
\Kl $\to \gamma \gamma$\ is known about 2.5\%.  To determine the
contribution when one or more of the photons is off-shell requires
determination of the form factor for the \Kggv vertex.  With that
and \BR{\Kl}{\mpmm} , which is now known to $\pm2.2$\%, we
can determine the short range contribution to \Kl $\to$ \mpmm, and
thereby determine \row.  {\bf This is the reason for our interest in
the form factor.}  After discussing experimental information from
KTeV about the form factor in four different modes, I will return
to a pressing theoretical issue in this scheme for constraining
\row.
\subsection{\Kl $\to$ \mpmm\epem}

The decay \Kl $\to$ \mpmm\epem\ will allow determination of the \Kggv
form factor directly from the mass spectrum of the dilepton pairs.
The signal appears as a set of four tracks in the
magnetic spectrometer, two of which point to clusters in the
calorimeter with energy within $\pm5\%$\ of the momentum measured in
the spectrometer, and two of which point to clusters of low energy
(we required $<3\gev$) and which extrapolate out to the muon counter
$\mu$3.  The major backgrounds are
\begin{itemize}
\item {\Kl $\to$ \mpmm$\gamma$, with $\gamma \to$ \epem\ through
       detector interaction.  This background is reduced by requiring
       that the tracks of the charged particles are well separated at
       the first drift chamber, and that the mass of the $e^\pm$\
       pair be over $3\mev$.  After selection requirements, this
       background is estimated to be 0.13 events.  This estimate uses
       a Monte Carlo simulation which is well-tuned to the data in
       regards to electromagnetic interactions in the vacuum window,
       where nearly all of the important interactions occur.}
\item {\Kl $\to \pi^+\pi^-\pi^0$, with $\pi^0 \to$ \epem$\gamma$,
       where both $\pi^\pm$\ fake $\mu^\pm$\ and the $\gamma$\ goes
       undetected.  As KTeV was designed with minimal detector
       mass in the the tracking system, we have limited ability
       to detect the kink which should in principle exist for the
       $\pi^\pm \to \mu^\pm$\ decays which make up the bulk of this
       misidentification category.  However, we can verify that the
       tracks form a good vertex in the decay region and that the
       track segments upstream of the magnet are close to the 
       segments downstream of the magnet at the magnet's bend plane.
       With the simulation, we estimate 0.03 events
       background from this source.  That number includes the
       contribution from the experimentally similar
       \Kl $\to \pi^+\pi^-$\epem.}
\item {Events where two simultaneous \Kl\ decays produced four charged
       particles appearing as an \epem\ pair and a \mpmm\ pair - for
       example, two \Kl $\to \pi^\pm e^\mp \nu$ decays with $\pi^\pm$\
       that appear as $\mu^\pm$.  This background level is estimated
       at 0.02 events using wrong-sign combinations from the data.}
\end{itemize}
KTeV has made the first observation of \Kl $\to$ \mpmm\epem, finding
38 events and determining
\BR{\Kl}{\mpmm\epem} \ =$(2.50 \pm0.41 \pm0.15) \times 10^{-9}$.  This
preliminary result is in good agreement with the VDM
model\cite{ref:oldie} and differs from both the
\omg{p^6} \ $\chi$PT prediction\cite{ref:Zhang} and Uy's
model.\cite{ref:Oy}  With this event sample, although we can make
statements about $CP$ violation as in Uy's model, we can not at this
time say anything about the form factor.  Over a hundred events are
expected with the inclusion of the 1999 data.
\subsection{\Kl $\to$ \epem\epem}

The decay \Kl $\to$ \epem\epem\ provides a much larger sample than
\Kl $\to$ \mpmm\epem\ to work with.  The signal is similar to the
\mpmm\epem\ signal, except of course for the particle identification.
The major backgrounds are
\begin{itemize}
\item {\Kl $\to$ \epem$\gamma$, with $\gamma \to$ \epem\ through
       detector interaction.  Similarly, there is the decay
       \Kl $\to \gamma\gamma$, with two $\gamma$\ conversions.  Again,
       track spacing requirements at the first drift chamber are
       required.  After selection requirements, this background
       is $3.0\pm0.3$\ events.}
\item {\Kl $\to \pi^\pm e^\mp \nu \gamma$, with pair conversion
       and a $\pi^\pm$\ misidentified as an $e^\pm$.  The primary 
       cause of this misidentification is from $\pi^\pm$\ which
       go down the beam hole in the calorimeter and have no associated
       cluster; to keep the signal acceptance high, these events are
       permitted. Thus, the rate of this background is largely
       controlled by the geometry of the detector, which is easily
       modeled in simulation.  This background is $0.5\pm0.5$\ events.}
\end{itemize}
KTeV's preliminary result is 
\BR{\Kl}{\epem\epem} \ =$(3.73 \pm0.18 \pm0.27) \times 10^{-8}$.  This
is in good agreement with expectations,\cite{ref:eeee} as are certain
angular distributions which show indirect $CP$ violation in this
mode.  With a sample of 436 events (before background subtraction), we
can begin to discuss the form factor.  However, the radiative
corrections for electronic decays need to be handled carefully and our
form factor analysis is currently being examined by the collaboration.
The branching ratio analysis is limited by systematic uncertainties and
is likely to stay that way.  The form factor analysis will probably
benefit from the improved statistics of the 1999 data.
\subsection{\Kl $\to$ \mpmm$\gamma$}

The decay \Kl $\to$ \mpmm$\gamma$ provides a still larger sample to work
with, albeit with one of the photons on-shell.  The signal appears as
two vertexable tracks of opposite sign, and a single calorimeter cluster
that is unassociated to any track and which combines with the tracks
to have a reconstructed mass within $8\mev$\ of the \Kl\ mass.  The
background is \Kl $\to \pi^\pm \mu^\mp \nu$, with a $\pi^\pm$\
misidentified as a $\mu^\pm$, and an accidentally coincident
calorimeter cluster not associated to any charged particle's track.
This cluster is often not a photon; often it is from some other
particle, and this background is reduced by requiring that the
cluster have a transverse shower profile consistent with that of
an electromagnetic shower.  These accidental "photons" are typically of
low energy, and this background is further reduced by requiring
$E_\gamma > 8\gev$.  Interpolating the reconstructed kaon mass
distribution from our data using the shape of this distribution from
the simulation, we find this background to be $222 \pm 15$\ events.
All other backgrounds added together are less than the uncertainty in
this estimate.

\begin{figure}
\vspace{7.3cm}
\includegraphics{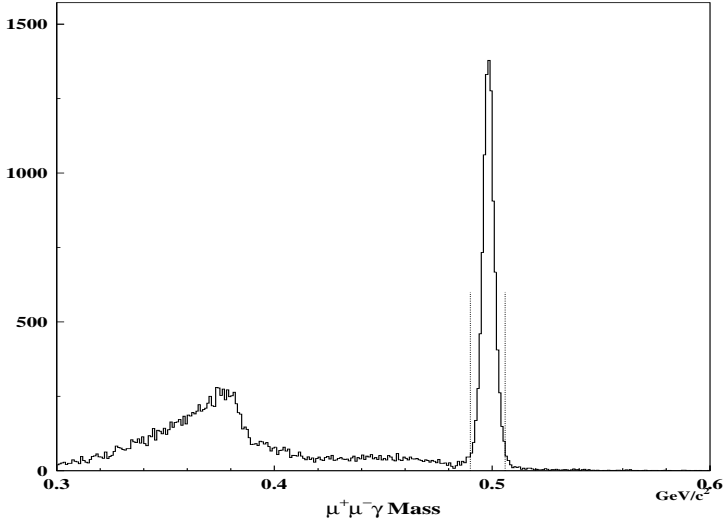}
\caption{\it
Reconstructed $m_{\mu\mu\gamma}$\ distribution after selection
criteria.  The feature near $380\mev$\ is from
\Kl $\to \pi^+ \pi^- \pi^0$; the \Kl $\to \pi^\pm \mu^\mp \nu$\
background dominates from $\sim 400$---$600\mev$, with a slight
enhancement from \Kl $\to \pi^\pm \mu^\mp \nu \gamma$\ at $450\mev$.
\label{fig:mmg_bump} }
\end{figure}

Figure \ref{fig:mmg_bump} shows the reconstructed kaon mass
distribution.\cite{ref:BQ}  There are 9327 events in the mass window
$490<m_{\mu\mu\gamma}<506\mev$, and our preliminary result is
\BR{\Kl}{\mpmm$\gamma$} \ =$(3.66 \pm0.04 \pm0.07) \times 10^{-7}$.
Existing predictions are roughly comparable, but depend greatly on the
assumed form factor.    The measurement of this decay channel is now
limited by systematic uncertainties that will not be reduced by
including the 1999 data.  With this sample of data in a
muonic mode, we can easily make reliable measurements of the form
factor.

There are two widely used models of the \Kggv\ form factor: that of
Bergstr\"{o}m, Mass\'{o} and Singer,\cite{ref:BMS} (BMS) and that of
D'Ambrosio, Isidori, and Portol\'{e}s\cite{ref:DIP} (DIP).  The 
first is a vector dominance model with one parameter, \akst, which
quantifies the relative contribution of the vector meson and
pseudoscalar diagrams in $K_L \to \gamma \gamma^{(*)}$\ decays.  To
apply the BMS form to decays with two virtual photons, we use the
product of two form factors for single $\gamma^*$s.  The second model
has two parameters, $\alpha_{\mathrm DIP}$\ and $\beta_{\mathrm DIP}$,
and has the properties of (a) being consistent with
\omg{p^6} \ $\chi$PT, (b) including the poles of vector resonances
of arbitrary residues, (c) having parameters that can be experimentally
determined in the low-$q^2$\ limit and (d) observing certain constraints
from QCD which apply to the high-$q^2$\ limit.

We determine the form factor parameters using the measured branching
ratio and the integrated form factors alone.  We also measure the form
factor parameters by fitting the $m_{\mu\mu}$\ distributions; the two
methods produce consistent results which we combine, yielding
\akst = -0.157\pma{0.025}{0.027} \ and
$\alpha_{\mathrm DIP} = -1.52 \pm0.09$.  Because one of the photons
is on-shell, sensitivity to $\beta_{\mathrm DIP}$\ is identically
zero.  (In \Kl $\to$ \epem\epem, sensitivity to
$\beta_{\mathrm DIP}$\ is practically zero because virtual photons
tend to materialize as low mass \epem\ pairs).  With these results in
hand, we repeat the analysis of DIP ({\it op.cit.}), and conclude that
\row$>-1.0$\ (BMS form factor) or \row$>-0.2$\ (DIP form factor). 
\subsection{\Kl $\to$ \epem$\gamma$}

This mode provides copious statistics: there will be \omg{10^5} \ events
in the 1997 data sample alone.  Obviously, systematic uncertainties need
to be well understood, and radiative corrections are critical.  The
analysis of the KTeV data for this mode is underway.
\subsection{Discussion of Constraints on \row}

The constraint that we are presently able to set on \row\ is not yet
as stringent as those than can be set by other means,\cite{ref:lims}
and it shows strong variation with the model used to extrapolate
from $\gamma^*\gamma$\ to $\gamma^*\gamma^*$\ decays.  The resolution
of this extrapolation problem should be possible with more data in
the \epem\epem\ and \mpmm\epem\ modes, but another hurdle lies beyond
this one.

To fully calculate the contribution from the diagrams in
Fig. \ref{fig:mm_long}, one needs to understand the form factor in the
high-$q^2$\ limit, but from \Kl\ decay data, one can only measure the
form factor up to $q^2$\aplt $\,m_K^2$.  There are three important 
recent papers addressing this issue: that of DIP ({\it op.cit.}),
that of Valencia\cite{ref:noway} and that of G\'{o}mez Dumm and
Pich.\cite{ref:Dumm}  It is beyond the scope of this talk to discuss
these papers in detail, but it should be noted that (a) Valencia seems
to come to a more pessimistic conclusion than the other authors and
(b) there does not seem to be a detailed calculation of what form
factors permit interesting limits on \row\ in the SM scenario or how
much theoretical uncertainty will be introduced into a measurement of
\row, should that be possible.  In any event, a better understanding of
the components of \Kl $\to$ \mpmm\ may well be of use in setting bounds
on possible New Physics scenarios.  It is also true that the power of
this technology will increase rapidly with better precision on 
\BR{\Kl}{\mpmm} \ and \BR{\Kl}{$\gamma\gamma$} ; specifically, one wants
the most precise value for
\BR{\Kl}{\mpmm} -($11.950 \time 10^{-6}$)\BR{\Kl}{$\gamma\gamma$} .
is .
\section{Modes Relevant for \ate}

\begin{figure}
\vspace{5.2cm}
\includegraphics{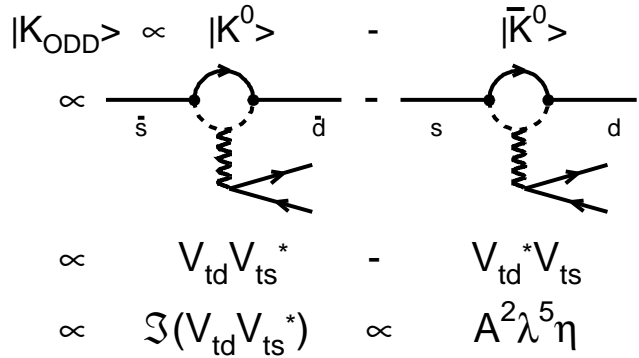}
\caption{\it
Interesting short distance contributions to
\Kl $\to \pi^0 \ell^+ \ell^-$.  There are also box diagrams with
similar amplitudes.
\label{fig:pill} }
\end{figure}

The decays \Kl $\to \pi^0$\lplm\ have been extensively studied for many
years;\cite{ref:lplm} the basic idea becomes evident after drawing the
diagrams of Fig. \ref{fig:pill}.  The real parts of the two components
of the \Kl\ decay amplitude cancel, leaving an amplitude proportional
to \ate.  In the case where the $\ell^\pm$\ is an $e^\pm$\ ($\mu^\pm$),
there are also amplitudes from indirectly $CP$ violating and
$CP$ conserving processes, and backgrounds from radiative (muonic)
Dalitz decay of the \Kl.  The neutrino case is very clean
theoretically, and has become the central topic of kaon physics
despite formidable experimental obstacles.  The KOPIO collaboration
at Brookhaven and the KaMI collaboration at Fermilab plan to measure
\BR{\Kl}{$\pi^0$\nunbr} .  The CKM collaboration at Fermilab
plans to measure \BR{\Kl}{$\pi^+$\nunbr} ; the charged mode is similar
except that the cancellation in Fig. \ref{fig:pill} does not occur.
Using numbers in reference\cite{ref:BnB2}, one obtains a constraint on
\mbox{(\ate)$^2$ +[(\row - $(5.40\pm0.66)]^2$}\
rather than upon $|$\ate$|$\ only.

With some oversimplification, we may compare the measurement of
CKM elements with semileptonic kaon decay to the measurement of CKM
elements with semileptonic $b$\ meson decay in the following way.  With
$b$\ mesons, we can extract leptonic coupling constants because we have
the HQET symmetry to help us understand the hadronic side; with kaons,
we can extract leptonic coupling constants because we have precise
experimental data to help us understand the hadronic side.
Expressions for \BR{\Kl}{$\pi^0$\nunbr} \ and 
\BR{\Kl}{$\pi^+$\nunbr} \ explicitly\cite{ref:BnB2} contain
\BR{$K^+$}{$\pi^0 e^+ \nu$} \ .

For these channels, the KTeV results presented in this talk have
either been published or been accepted for publication\cite{ref:KTeV}
since the conference.  All of the searches resulted in limits which are
orders of magnitude more stringent than previously available limits,
although sensitivities are not yet good enough to see Standard Model
physics.  For both $\pi^0$\epem\ and $\pi^0$\mpmm, backgrounds have
begun to limit the experimental reach to
$\sqrt{\mathrm {dataset~size}}$.  The \Kl $\to \pi^0$\nunbr\ triggers
were disabled for the 1999 data.
\section{Modes relevant for lepton flavor violation}

A detector capable of detecting \Kl $\to$ \mpmm\epem, $\pi^0$\epem\
and  $\pi^0$\mpmm\ is also a detector capable of detecting lepton
flavor violation.
\subsection{\Kl $\to$ $\mu^+\mu^+e^-e^-$}

There were no wrong sign combination events from the
\Kl $\to$ \mpmm\epem\ analysis.  From this we obtain a preliminary
result, \BR{\Kl}{$\mu^+\mu^+e^-e^-$} \ $< 1.36 \times 10^{-10}$\ at the
90\% C.L.  We simulated signal events with a flat phase-space
distribution to calculate the acceptance.
\subsection{\Kl $\to \pi^0 \mu^\pm e^\mp$}

Searches for this decay complement searches for
\Kl $\to$ $\mu^\pm e^\mp$\ by being sensitive to new scalar or vector
interactions; the two body decay would be the result of a pseudoscalar
interaction.  The major backgrounds are
\begin{itemize}
\item {\Kl $\to \pi^\pm e^\mp \nu$, with the $\pi^\pm$\ faking
       a $\mu^\pm$, and with two accidental "photons".  In addition to
       the types of cuts used in the modes above, this mode may be
       suppressed by requiring that the momentum transverse to the
       \Kl\ line of flight is small.  This is the dominant background,
       contributing $0.61 \pm0.56$\ events.}
\item {\Kl $\to \pi^\pm e^\mp \nu \gamma$, with the $\pi^\pm$\ faking
       a $\mu^\pm$, and with only one accidental "photon".  The
       branching ratio is small, and by requiring that the reconstructed
       neutrino momentum squared in the \Kl\ frame is non-positive,
       this background can be reduced to less than 0.0054 events.}
\item {\Kl $\to \pi^+\pi^-\pi^0$, with one $\pi^\pm$\ faking
       a $\mu^\pm$\ and the other appearing as an $e^\pm$.  The
       particle ID power of the calorimeter is augmented with the
       TRDs for this analysis, and this background is negligible
       in the signal region.  It does however appear at low
       reconstructed \Kl\ mass.}
\end{itemize}
We found two candidate events, and given the level of uncertainty in 
our present background estimates, have chosen to not subtract background
for our preliminary result.  We find 
\BR{\Kl}{$\pi^0 \mu^\pm e^\mp$} \ $< 4.4 \times 10^{-10}$.
In terms of a model based on $SU(n)$\ family symmetry\cite{ref:CnH} with
couplings equal to $g_{\mathrm EW}$, this corresponds to a vector boson
of $>44 \,\mathrm {TeV}$.  In comparing our results to those of
Brookhaven E865, one must allow for the ratio of \Kl\ and $K^+$\
lifetimes and for the fact that E865 is far more sensitive to
$\pi^+ \mu^+ e^-$\ than to $\pi^+ \mu^- e^+$, while KTeV can see both
charge combinations well.
\section{Appendix: Description of the KTeV Detector}
\label{sec:det}
In the 1997 data taking, an 800 GeV proton beam, with typically
$3.5\times 10^{12}$\ protons per $\sim$20 second Tevatron spill every
minute, was targeted at a vertical angle of $4.8\mrad$\ on a
1.1 interaction length ($300\mm$) BeO target.  Photons were converted
by $76\mm$\ of lead immediately downstream of the target.  Charged
particles were then removed with magnetic sweeping.  Collimators
defined two $0.25\,\mu\mathrm {sr}$\ beams that entered the KTeV
apparatus 94 meters downstream of the target.  About
$14 \times 10^7$\ neutral kaons per second entered the 65 meter vacuum
($\sim10^{-6}$\ Torr) decay region which extended to the first drift
chamber.  The spectrometer consisted of a dipole magnet and four drift
chambers.  The drift chambers ranged from $1.28 \times 1.28\,m^2$\ to
$1.77 \times 1.77\,m^2$\ in size, and had $\sim100\um$\ position
resolution in both horizontal and vertical directions.  Helium filled
bags occupied the spaces between the drift chambers; the Kevlar
reinforced mylar window which sealed off the vacuum region was
just upstream of the first drift chamber, and converted 
$(2.74 \pm0.11) \times 10^{-3}$\ of the incident photons into 
$e^+e^-$\ pairs.  The magnet's field was uniform to $\sim1$\% and
was mapped to $\sim1$\ part in $10^{4}$\ over the volume of the
pole gap; it imparted a $200\mev$\ horizontal momentum kick.
The spectrometer had a momentum resolution of
$\sigma(P)/P = 0.38\% \oplus 0.016\%P$, where P is in $\gev$.  The
electromagnetic calorimeter consisted of 3100 pure CsI crystals.
Each crystal was $500\mm$\ (27 radiation lengths, 1.4 interaction
lengths) long.  Crystals in the central $1.2 \times 1.2\,m^2$\ section
of the calorimeter had a cross-sectional area of $25 \times 25\mm^2$,
and those in the outer region (out to $1.9 \times 1.9 m^2$) had a
$50 \times 50\mm^2$\ area.  The calorimeter's energy resolution for
photons was $\sigma(E)/E = 0.45\% \oplus 2\% / \sqrt{E}$, where E is in
$\gev$, and its position resolution was $\sim1\mm$.  Additional
$e^\pm / \pi^\pm$\ separation was provided with eight transition
radiation detectors (TRDs) located behind the spectrometer.
The TRDs used polypropylene felt for radiators and  80/20\% Xenon/CO2
filled MWPC volumes to detect transition radiation.  They provided
pion rejection factors at 90\% electron acceptance which varied between
200::1 and 300::1 through the data taking period.  Nine photon veto
assemblies (lead scintillator sandwiches of 16 radiation length
thickness) detected particles leaving the fiducial volume.  Two
scintillator hodoscopes in front of the calorimeter were used to
trigger on charged particles.  The hodoscopes and the calorimeter had
two holes ($150 \times 150\mm$\ at the calorimeter) to let the neutral
beams pass through without interaction.  The muon filter, located
behind the calorimeter, was constructed of a $100\mm$\ thick lead wall
followed by three steel walls of 1.04, 3.04, and $1.03\,m$\ thickness.
Scintillator planes with $150\mm$\ segmentation in both horizontal and
vertical directions ($\mu3$) were located after the third steel wall.
The segmentation was comparable to the multiple scattering angle of
$10\gev$\ muons at $\mu3$.  The pion punch-through probability,
including decays downstream of the calorimeter, was determined as a
function of momentum from $K_{e3}$\ data, and is on the order of a few
times $10^{-3}$.  The data acquisition system reconstructed on the
order of $10^5$\ events of 7 kbyte size per Tevatron spill online,
and the results were used to filter the data.

In 1999, the Tevatron spill was extended to $\sim$40 seconds and the
spectrometer magnet imparted a $150\mev$\ horizontal momentum kick.
\end{document}